\documentclass{ws-procs9x6}

\newcommand {\beq}{\begin{eqnarray}}
\newcommand {\eeq}{\end{eqnarray}}
\newcommand {\non}{\nonumber\\}

\newcommand {\1}[1]{\frac{1}{#1}}

\newcommand {\ph}{\varphi}

\newcommand {\sig}{\sigma}

\newcommand {\Sig}{\Sigma}

\newcommand {\del}{\partial}
\newcommand {\dagg}{^{\dagger}}

\newcommand {\tr}{{\rm tr}\,}

\newcommand{\hs}[1]{\hspace{#1 mm}}

\begin{document}
\title{Domain Walls with Massive Hypermultiplets
\footnote{\uppercase{T}alk presented by \uppercase{M.N.}
at {\it \uppercase{SUSY} 2003:
\uppercase{S}upersymmetry in the \uppercase{D}esert}\/, 
held at the \uppercase{U}niversity of \uppercase{A}rizona,
\uppercase{T}ucson, \uppercase{AZ}, \uppercase{J}une 5-10, 2003.
\uppercase{T}o appear in the \uppercase{P}roceedings.}}
%%  PLEASE USE THE ABOVE FOOTNOTE FOR ALL ARXIV POSTINGS,
%%   (SUBSTITUTING SPEAKER NAME, OF COURSE;  NO SPEAKER NAME NEEDED
%%      FOR SINGLE-AUTHOR CONTRIBUTIONS). 
%%  NOT NECESSARY TO HAVE FOOTNOTE FOR VERSIONS THAT ARE SENT TO 
%%  SUSY 2003 CONFERENCE FOR PUBLICATION

\author{Masato Arai}
\address{Institute of Physics, AS CR, 
182 21, Praha 8, Czech Republic}  

\author{Muneto Nitta\footnote{
\uppercase{C}urrent address : \uppercase{T}okyo 
\uppercase{I}nstitute of \uppercase{T}echnology.}}
\address{Department of Physics, Purdue University, 
West Lafayette, IN 47907-1396, USA}

\author{Norisuke Sakai}
\address{Department of Physics, Tokyo Institute of 
Technology, Tokyo 152-8551, JAPAN}  


\maketitle

\abstracts{
Massive hypermultiplets admit degenerate discrete vacua 
only if they form nonlinear sigma models 
or have gauge interaction.
We discuss BPS domain walls in these theories.
This talk is based on the original papers.\cite{ANS,ANNS}
}

\section{Introduction}
Domain walls play a central role 
in a subject of recent interest, 
the brane-world scenario.\cite{brane} 
In particular, 
BPS domain walls are investigated in detail 
in $D=4$, ${\mathcal N}=1$ SUSY theories.\cite{NNS}
We should consider hypermultiplets 
to realize $D=4$, ${\mathcal N}=1$ SUSY 
theory on the world-volume.
In order to have a potential with 
discrete degenerate vacua,
they must have gauge interaction,
or nonlinearity of kinetic term, 
namely nonlinear sigma models (NLSM).
The latter case can be obtained from the former case 
in the strong gauge coupling limit. 
A lot of interesting solitons have been discussed in 
hypermultiplets with\cite{soliton} or 
without\cite{soliton2} taking 
this limit.

In this talk, the hyper-K\"ahler (HK) quotient method\cite{LR} 
is generalized to the massive models 
and the BPS domain wall in the simplest case is given. 
Keeping essential properties of eight supercharges,
we discuss a simpler and familiar case of 
$D=4$, ${\mathcal N}=2$ SUSY theories 
in the ${\mathcal N}=1$ superspace formalism. 
Brief reviews can be found in reports.\cite{ANS2}
The Harmonic superspace formalism is discussed in 
the original paper.\cite{ANS}

%%%%%%%%%%%%%%%%%%%%%%%%%%%%%%%%%%%%%%%%%%%%

\section{Walls in Chiral Multiplets}

We discuss BPS walls in 
${\mathcal N}=1$ SUSY theories with 
chiral superfields, 
whose bosonic Lagrangian
is ${\mathcal L}_{\rm boson}  
 = - g_{ij^*} \del_{\mu}\phi^i \del^{\mu}\phi^{*j}  
   - g^{ij^*} \del_i W \del_{j^*} W^*$
with $g_{ij^*}$ 
the K\"ahler metric. 
Assuming a domain wall configuration perpendicular to $z$,
its energy density per unit area in the $x$-$y$ plane is given by
\beq
 E &=& \int d z  
 ( g_{ij^*} \del_z \phi^i \del_z \phi^{*j}
   + g^{ij^*} \del_i W \del_{j^*} W^* ) \non
 &=& \int d z  
       |\del_z \phi^i - e^{i\alpha} g^{ik^*} \del_{k^*} W^*|^2
   + \int d z ( e^{i\alpha} \del_z \phi^i \del_i W  + {\rm c.c.} )\non
 &\geq& \int d z ( \del_z \phi^i \del_i W   + {\rm conj.} ) 
 = 2 {\rm Re} (e^{i\alpha} \Delta W) \;
\eeq
with the norm defined by $|V^i|^2 \equiv g_{ij^*}V^iV^{*j}$, 
$\Delta W \equiv W|_{z=\infty} - W|_{z=-\infty}$ 
and $\alpha$ an arbitrary real constant.
Since we obtain the best bound at
$e^{-i\alpha} = \Delta W/|\Delta W|$, 
we derive the BPS bound 
$E \geq 2 |\Delta W|$ saturated by 
solutions of the BPS equation 
$\del_z \phi^i = e^{-i\alpha} g^{ij^*} \del_{j^*} W^*$. 
Since the SUSY transformation on the fermion $\psi^i$ 
is calculated as
$\delta_{\epsilon} \psi^i 
= \sqrt{2} (i \sig^{\mu} \bar{\epsilon} \del_{\mu} \phi^i 
+ \epsilon F^i ) 
= \sqrt {2} (i \sig^{z} \bar{\epsilon} e^{-i\alpha} - \epsilon) 
g^{ij^*}\del_{j^*}W^*$, 
two SUSYs satisfying 
$i e^{-i\alpha} \sig^{z} \bar{\epsilon}  = \epsilon$ 
are preserved.

%%%%%%%%%%%%%%%%%%%%%%%%%%%%%%%%%%%%%%%%%%%%%%%%%%%
\section{Walls in Hypermultiplets}

Let $(\Phi,\Psi)$ be ${\mathcal N}=2$ hypermultiplets 
with $\Phi$ and $\Psi$ being 
$N \times M$ and $M \times N$
matrix chiral superfields. 
To obtain nontrivial vacua we need an $U(M)$ gauge symmetry
introducing ${\mathcal N}=2$ vector multiplets 
$(V,\Sigma)$ with $V$ an $M \times M$  matrix vector superfield
and $\Sigma$ an $M \times M$ matrix chiral superfield. 
We work out for the $U(M)$ gauge group in which 
$U(1)$ part is essential to obtain discrete vacua.
We consider the Higgs branch of the theory 
taking the strong coupling limit $g\to \infty$ 
of gauge interactions, which eliminates 
the kinetic terms for $V$ and $\Sig$.
The gauge invariant Lagrangian is given by
\beq
&& {\mathcal L} = \int d^4 \theta
 \left[ \tr (\Phi\dagg\Phi e^V )  
 + \tr (\Psi\Psi\dagg e^{-V}) - c \, \tr V \right]  \non
&&\hs{5} + \left[ \int d^2\theta \,
  \big\{
  \tr \{ \Sigma (\Psi \Phi - b {\bf 1}_M) \} 
    +  {\sum_{a=1}^{N-1} m_a \tr (\Psi H_a \Phi)} \big\} 
         + {\rm c.c.}\right] ,
\label{linear}
\eeq
with $b \in {\bf C}$ and $c \in {\bf R}$ 
constituting a triplet of the Fayet-Iliopoulos parameters, 
$m_a$ complex mass
and $H_a$ Cartan generators of $SU(N)$.\footnote{
Flavor symmetry   
$\Phi \to \Phi' = g \Phi$,
$\Psi \to \Psi' = \Psi g^{-1}$ with 
$g \in SU(N)$ in the massless limit $m_a =0$
is explicitly broken by the mass to 
its Cartan $U(1)^{N-1}$ generated by $H_a$.
}
Eliminating superfields $V$ and $\Sig$ using 
their algebraic equations of motion, 
we obtain the Lagrangian in terms of independent superfields,
in which the K\"ahler potential is
\beq
 &&K = c\, \tr \sqrt{{\bf 1}_M + {4\over c^2} \Phi\dagg\Phi \Psi\Psi\dagg} 
%  \non && \hs{10} 
   - c\, \tr \log \left( {\bf 1}_M 
    + \sqrt{{\bf 1}_M + {4\over c^2} \Phi\dagg\Phi \Psi\Psi\dagg}\right) \non
 && \hs{10}
    + c\, \tr \log \Phi\dagg\Phi  \;, \label{kahler}
\eeq
with a gauge fixing\footnote{
We discuss the $b \neq 0$ case here. 
The $b=0$ case must be discussed independently.\cite{ANS}
}
\beq
 \Phi = \begin{pmatrix}
          {\bf 1}_M \cr \ph
        \end{pmatrix}  
        Q \;, \hs{5} 
 \Psi = Q ({\bf 1}_M, \psi) \;, \hs{5}
 Q = \sqrt b ({\bf 1}_M + \psi\ph)^{-\1{2}} \;, 
 \label{fixing2}
\eeq
\if0 %%%
\beq
 \Phi^T = ({\bf 1}_M, \ph^T) \;, \hs{5} 
 \Psi = (- \psi\ph, \psi) \;, \label{fixing1}
\eeq
\fi %%%
with $\ph$ ($\psi$) an $(N-M) \times M$ [$M \times (N-M)$] matrix 
chiral superfield, 
and the superpotential is
\beq
 W = b \sum_a m_a \tr \left[
    H_a \begin{pmatrix}
          {\bf 1}_M \cr \ph
        \end{pmatrix} 
    ({\bf 1}_M + \psi\ph)^{-1}
    ({\bf 1}_M, \psi)  
   \right] \;. \label{superpot2}
\eeq
\if0 %%%
\beq
&& W 
= - \sum_{\alpha =1}^M \sum_{i=1}^{N-M} 
     M_{\alpha i} \ph_{i \alpha} \psi_{\alpha i} \;, \non
 && M_{\alpha i} \equiv 
   \sqrt{i+M-1 \over i+M} m_{i+M-1} 
 - \sqrt{\alpha-1\over \alpha} m_{\alpha - 1} 
 + \sum_{a = \alpha}^{i+M-1} {m_a \over \sqrt {a(a+1)}}\;,
\eeq
\fi %%%
This is the massive extension of the HK NLSM on 
the cotangent bundle over the Grassmann manifold, 
$T^* G_{N,M}$, found by Lindstr\"om and Ro\v{c}ek.\cite{LR}
This model contains 
${}_N C_M = N!/M! (N-M)!$ discrete degenerate vacua 
corresponding to independent gauge fixing 
conditions (\ref{fixing2}).\cite{ANS}

%%%%%%%%%%%%%%%%%%%%%%%%%%%%%%%%%%%%%%%%%%%%%%%%%%%
%\section{Domain Wall Solutions}
%As seen in Section 2, the tension of the BPS domain wall 
% is given by superpotential. 
%This fact implies that 
%We should take $b \neq 0$ for simplicity. 
% in ${\mathcal N}=1$ superfields. 
The $M=1$ case of $U(1)$ gauge symmetry 
reduces to $T^* {\bf C}P^{N-1}$ with the superpotential, 
which admits $N$ parallel domain walls.
Moreover if we take $N=2$ and $M=1$, the target space 
$T^*{\bf C}P^1$ is the Eguchi-Hanson space with the superpotential 
$W = b {\mu \over 1 + \ph \psi}$ ($\mu \equiv m_1$).
The BPS equation in ${\mathcal N}=1$ superfields 
can be solved to give\cite{ANNS}
\beq
 \ph = \psi^* = e^{|\mu| (z - z_0)} e^{i \delta}\;,    
 \label{solution} 
\eeq
where $z_0$ and $\delta$ are integral constants. 
They correspond to zero modes 
arising from spontaneously broken 
translational invariance perpendicular 
to domain wall configuration and 
$U(1)$ isometry $\sig_3$ in the {\it internal} space.

%%%%%%%%%%%%%%%%%%%%%%%%%%%%%%%%%%%%%%%%%%%%%%%%%%%%%%%%%%%%%
%                                                           %
% You may repeat \section{SECTION N-th HEADING TYPE HERE}   %
%                                                           %
% Do start a subsection or sub-subsection, do this:-        %
%                                                           %
%   \subsection{SUBSECTION HEADING TYPE HERE}               %
%                                                           %
%   \subsubsection{SUBSUBSECTION HEADING TYPE HERE}         %
%                                                           %
% instead of the above                                      %
%                                                           %
%%%%%%%%%%%%%%%%%%%%%%%%%%%%%%%%%%%%%%%%%%%%%%%%%%%%%%%%%%%%%

\if0
\section{Conclusions}
We have constructed massive HK NLSM 
on the cotangent bundle over $G_{N,M}$ in 
${\mathcal N}=1$ superfields,
which is the massive extension of Lindstr\"om and Ro\v{c}ek.
This model contains ${}_N C_M = N!/M! (N-M)!$ 
discrete degenerate vacua. 
A BPS Domain wall solution in the simplest $T^* {\bf C}P^1$ 
has been given. 

Constructing domain walls in non-Abelian gauge group 
remains as an interesting future work. 
Turning on the gauge coupling does not change vacua. 
BPS walls should also be similar as shown 
in the $M=1$ case.\cite{TL,KSIOS}
Coupling to supergravity is possible as  
in the $M=1$ case.\cite{sugra}

\fi

\section*{Acknowledgements}
We would like to thank
Masashi Naganuma for discussions in the early stage 
of this work. 
M.~N. is grateful to the organizers in 
SUSY2003.
This work is supported in part by Grant-in-Aid 
 for Scientific Research from the Japan Ministry 
 of Education, Science and Culture  13640269  (NS). 
The work of M.~N. was supported by the U.~S. Department
 of Energy under grant DE-FG02-91ER40681 (Task B).

%%%%%%%%%%%%%%%%%%%%%%%%%%%%%%%%%%%%%%%%%%%%%%%%%%%%%%%%%%%%%%%%%%%%%%%

\end{document}